# WAVE PARTICLE DUALITY, THE OBSERVER AND RETROCAUSALITY

Ashok Narasimhan[a,b] and Menas C. Kafatos[c,d,e]

[a]*Cofounder and President of the Nalanda Institute for Consciousness Study and Research (NICSaR), San Francisco, CA*
[b]*Board of Trustees of the California Institute of Integral Studies (CIIS), San Francisco, CA*
[c]*Fletcher Jones Endowed Professor of Computational Physics, Chapman University, Orange, CA*
[d]*Outstanding Visiting Professor, Korea University, Seoul, Korea*
[e]*Affiliated Researcher, National Observatory of Athens, Greece*

**Abstract.** We approach wave particle duality, the role of the observer and implications on Retrocausality, by starting with the results of a well verified quantum experiment. We analyze how some current theoretical approaches interpret these results. We then provide an alternative theoretical framework that is consistent with the observations and in many ways simpler than usual attempts to account for retrocausality, involving a non-local conscious Observer.

## THE WAVE PARTICLE DUALITY PROBLEM AND CURRENT APPROACHES

Here we examine current approaches to the well-known wave particle duality issues.

### Simple Double Slit Experiments

In the basic double slit experiment, a beam of light (usually from a laser) is directed perpendicularly towards a wall pierced by two parallel slit apertures. If a detection screen is put on the other side of the double slit wall, a pattern of light and dark fringes will be observed, a pattern that is called an interference pattern. By decreasing the brightness of the source sufficiently, individual particles that form the interference pattern are detectable [1]. The emergence of an interference pattern suggests that each particle passing through the slits interferes with itself, and that, therefore, in some sense the particles are going through both slits at once [2]: This is an idea that contradicts our everyday experience of discrete objects.

A well-known thought experiment, which played a vital role in the history of quantum mechanics (for example, see the discussion on Einstein's version of this experiment), demonstrated that if particle detectors are positioned at the slits, showing through which slit a photon goes, the interference pattern will disappear [3]. This which-way experiment illustrates the complementarity principle that photons can behave as either particles or waves, but not both at the same time [4][5][6].However, technically feasible realizations of this experiment were not proposed until the 1970s [7].

Which-path information and the visibility of interference fringes are hence complementary quantities. In the double-slit experiment, conventional wisdom held that observing the particles inevitably disturbed them enough to destroy the interference pattern as a result of the Heisenberg uncertainty principle.

### Quantum Eraser Experiments

However, in 1982, Scully and Drühl found a loophole around this interpretation [8]. They proposed a "quantum eraser" to obtain which-path information without scattering the particles or otherwise introducing uncontrolled phase factors to them. Rather than attempting to observe which photon was entering each slit (thus disturbing them), they proposed to "mark" them with information that, in principle at least, would allow the photons to be distinguished after passing through the slits. Lest there be any misunderstanding, the interference pattern does disappear when the photons are so marked. However, the interference pattern reappears if the which-path information is further manipulated after the marked photons have passed through the double slits to obscure the which-path markings. Since 1982, multiple experiments have demonstrated the validity of the so-called quantum "eraser"[9][10][11].

Versions of the quantum eraser using entangled photons, however, are intrinsically non-classical. Because of that, in order to avoid any possible ambiguity concerning the quantum versus classical interpretation, most experimenters have opted to use non-classical entangled-photon light sources to demonstrate quantum erasers with no classical analog.

Furthermore, use of entangled photons enables the design and implementation of versions of the quantum eraser that are impossible to achieve with single-photon interference, such as the delayed choice quantum eraser which is the topic of this paper.

The delayed choice quantum eraser experiment investigates a paradox. If a photon manifests itself as though it had come by a single path to the detector, then "common sense" (which Wheeler and others challenge) says it must have entered the double-slit device as a particle. If a photon manifests itself as though it had arrived by two indistinguishable paths, then it must have entered the double-slit device as a wave. If the experimental apparatus is changed while the photon is in mid‑flight, then the photon should reverse its original "decision" as to whether to behave as a wave or as a particle. Wheeler pointed out that when these assumptions are applied to a device of interstellar dimensions, a last-minute decision made on earth on how to observe a photon could alter a decision made millions or even billions of years ago

## SUMMARY OF CURRENT THINKING

In the Copenhagen Interpretation proposed initially by N. Bohr, W. Heisenberg, and further developed by them, including W. Pauli, M. Born and others [4][5], the role of observation plays a fundamental role in the so-called collapse of the wave function.

### Act of Observation

Here we summarize the traditional way of understanding the act of observation and the role of the observer (this is often referred to as the Orthodox view as H.P. Stapp and others refer to—see below):

A human being decides to observe the properties of a quantum system at a particular point in time and space, setting up appropriate experiment, as for example to determine the direction of the spin of a particle. The consciousness that plans, executes and subsequently makes the observation, is within this human being, in the same space-time coordinates.

We will term this the *Local conscious observer* (Lco). Lco sets up the apparatus needed to conduct the observation or 'probing action' to ask a question of Nature. Nature responds with its answer at $T_1$. This answer is recorded by the measuring apparatus at $T_1$. Lco then observes the recording of the answer at time $T_2$ which is $> T_1$. The act of observation causes the particular form of Nature's answer – e.g. particle or interference pattern. Here we note that this is based on the concept of a conscious observer using a traditional measuring apparatus to measure Nature's answer, all occurring at time $T_1$. It also raises the question of whether the act of observation disturbs the system being observed such that the results cannot be trusted to accurately represent its original state e.g. the interference pattern gets disturbed enough by the act of observation (Heisenberg's Uncertainty Principle) that it collapses to a particle.

## OUR APPROACH

Our approach is to start with verified experimental evidence and then construct the theory to explain those results. We will do this by eliminating the variables involved in a direct local measurement, with the potential for disturbing the system being observed, by using the delayed quantum eraser experiment described in detail in Kim *et al*. [12]. The set-up is illustrated in Fig 1.

An argon laser generates individual photons that pass through a double slit apparatus (vertical black line in the upper left hand corner of the diagram). An individual photon goes through one (or both) of the two slits. In the illustration, the photon paths are color-coded as red or light blue lines to indicate which slit the photon came through (red indicates slit A, light blue indicates slit B). After the slits, spontaneous parametric down conversion (SPDC) is used to prepare an entangled two-photon state.

One of these photons, referred to as the 'signal proton' (red and blue lines going upwards from the prism) goes to the detector $D_0$. The other entangled photons, referred to as the 'idler photons' (red and blue lines going downwards from the prism) pass through the arrangement of a prism, beam splitters and mirrors which direct them to the detectors labeled $D_1$, $D_2$, $D_3$, $D_4$. The arrangement ensures that if the idler photon is detected:

- at $D_3$ it can only have come from B ('which path' information is available)
- at $D_4$, it can only have from A (which 'path' information is available)
- at $D_1$ or $D_2$, it might have come from either A or B ( 'no which path' information is available).

The optical length from the slit to $D_1$, $D_2$, $D_3$ and $D_4$ is 2.5m longer than the length from the slit to $D_0$. This results in the information from the idler photon being available 8 ns later than the signal photon. Consequently, the detection of the idler photons provides delayed 'which path' information. $D_0$ is scanned along its x-axis by a stepper motor and using the coincidence counter, it is possible to record only events where both the signal and idler photons were detected, after compensating for the 8 ns delay. The plot with the position of $D_0$ on the x axis and the joint detection rates between $D_0$ and $D_1$, $D_2$, $D_3$ and $D_4$, resulted in charts (see [12]) that showed that:

When the experimenters looked at the signal photons whose entangled idlers were detected at $D_1$ or $D_2$ (these were the detectors where no which-path information was available), they detected interference patterns (wave aspect). When they looked at signal photons whose entangled idlers were detected at D3 or D4, they detected simple diffraction patterns, with no interference (particle aspect).

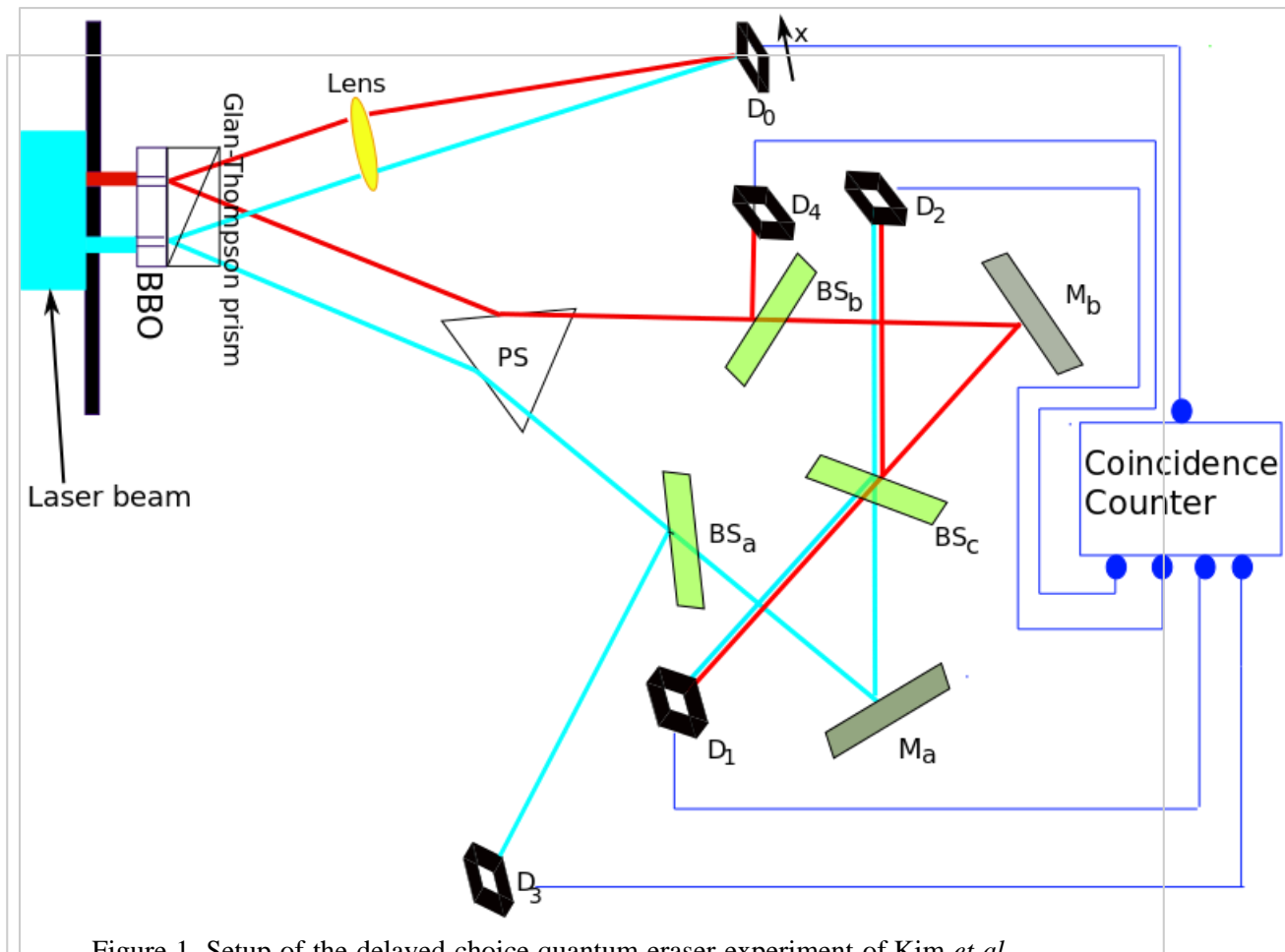

Figure 1. Setup of the delayed choice quantum eraser experiment of Kim *et al.* Detector D0 is movable (adopted from https://en.wikipedia.org/wiki/Delayed_choice_quantum_eraser)

What this showed was that when the signal photons were matched with the behavior of their entangled photons, the signal photons demonstrated interference or simple diffraction patterns depending on whether their entangled photons had, after an 8 ns delay, retained or lost (erased) the which-path information.

## Significance of the Results

Unlike the situation in the classic double-slit experiment, the choice of whether to preserve or erase the which-path information of the idler was not made until 8 ns after the position of the signal photon had already been measured by $D_0$. In other words, even though an idler photon is not observed until long after its entangled signal photon arrives at $D_0$ due to the shorter optical path for the latter, interference at $D_0$ is determined by whether a signal photon's entangled idler photon is detected at a detector that preserves its which-path information ($D_3$ or $D_4$); or at a detector that erases its which-path information ($D_1$ or $D_2$).

Now let us reflect on what Kim *et al*. experiment clearly demonstrates:

a) Setup

    The observer sets up the measuring apparatus at time $T_1$.

    This measuring apparatus does not require the ongoing active participation of asking probing questions of Nature by the observer.

    Instead, the apparatus uses the 50 % randomness of reflection vs transmission to generate the different alternatives.

    The apparatus makes a recording of an event concerning a particular signal photon at time $T_2$ at $D_0$.

    It makes two other recordings of the events of that photon's entangled twin, both at time $T_3$.

    One, when the 'which path' information is available ($D_3$, $D_4$).

    One, when this information is not available ($D_1$, $D_2$).

    The experiment is set up such that $T_2 - T_3 = 8$ ns.

    The experiment uses a coincidence counter to link each signal photon with its idler photon twin, after adjusting for the 8 ns delay.

    At time $T_4$, the human observe returns, then observes the recordings at $D_0$ and $D_1$ through $D_4$.

b) Understanding the results

We now see that it is the availability of 'which path' *information*, not the *act of measurement*, that triggers wave function collapse and determines the outcome – interference patterns or not. To put it another way, there is no 'probing action' or 'asking the question' that is required to trigger wave function collapse, WFC (see discussion below). Similarly, there is no 'Nature responding to the question posed' that is required to trigger WFC.

    The initial 'act of measurement' of the signal photon happens at $T_0$.

    This always results in no interference pattern, since there is always which-path information, at $T_0$.

    The measurement of the idler photon happens at $T_1$ and $T_1$ is 8 ns after $T_0$.

We now introduce two new terms here, 'initial signal photon' (ISP); this refers to the detection of the signal photon at $D_0$ at $T_0$; and 'modulated signal photon' (MSP); this refers to the 'modulation' of the signal photon when mapped to the detection at $D_x$ of its idler photon pair at $T_1$.

    At $T_1$, when the signal photons at $D_0$ are matched with the corresponding idler photons, the signal photons corresponding to the idler photons at $D_1$ and $D_2$ always show interference pattern, since 'the which' path information is 'erased'.

    Similarly, at $T_1$, the signal photons corresponding to the idler at $D_3$ and $D_4$ show no interference patterns.

    Note: this happens only when the information available at $T_1$ is obtained and superimposed on information available at $T_0$.

    The observer (Lco) actually observes the end results only at $T_2$, which can be minutes or hours after $T_1$.

    Therefore, the observer is not *observing* or *measuring* anything at $T_0$ or $T_1$ and is consequently temporally disassociated from the time when the measurement is made or the 'question asked of Nature' or 'Nature's response received'.

    What happens at $T_1$ is that the *information available overrides* the *information available* at $T_0$ and is then

observable by Lco at $T_2$.

We now discuss the implications of all this in the rest of this paper.

## Implications

There are several implications in this advanced delayed-choice, quantum eraser experiment, which were not available in previous versions and certainly not in normal double-slit experiments. Here we list them:

1. It is the availability of information at that 'point in time' that determines the end result
2. It is not the conscious act of observation by the Lco
3. The natural question then arises: Available to whom?
4. Is the 'unmarking, or scrambling' of information still information?
5. If it is still information, it is not recognized as such after it is unmarked/scrambled.
6. So that implies that the information has to 'be recognized or understood or be structured in some way' to be valid in causing the 'collapse of wave function'.
7. And the scrambled/unmarked information is 'seen' as random.
8. Therefore, it is not just information, but *structured* or *organized information* that provides the trigger.
9. And that implies that there is a 'Conscious Observer'*,* who is able to evaluate and interpret and differentiate the structured information content from 'random information'.
10. And since this Conscious Observer is able to interpret in a time independent way (e.g. $T_0$, $T_1$), the Conscious Observer has to be *outside space-time*, i.e., *non-local*.
11. To make sure everything is understood properly, we provide here other ways, step by step to examine the non-locality of the Conscious Observer. The arguments go as follows:

    The signal photon SP1 arrives at detector $D_0$ at $T_0$.
    The idler photon IP1 corresponding to this signal photon arrives at $D_3$ at $T_1$ ($T_1$ is 8 ns after $T_0$)
    Since which-path information is available to the Observer, it is reflected in the distribution pattern of the signal photon SP1, which is consistent with a general diffraction pattern (i.e. particle)
    This recording of SP1 is done at time $T_0$ and yet it reflects the information available at time $T_1$
    Similarly, signal photon SP2 arrives at $D_0$ at $T_2$
    Its idler photon IP2 arrives at $D_1$ at $T_3$ ($T_3$ is 8 ns after $T_2$)
    Since which-path information is NOT available for this to the Observer, it is reflected in the distribution pattern of the signal photon SP1, which is seen as an interference pattern.
    This recording of SP2 is done at time $T_2$, and yet it reflects the information available at time $T_3$, etc.

    This means that:

    The recording of the distribution pattern of SP1 was already set at $T_0$
    The distribution pattern as a 'particle' was based on the Observer having which-path information about its idler photon IP1, which information was available only at $T_1$
    Therefore, the only way this can happen is that the Observer was outside space-time
    Which also means that the Observer is non-local
    And this is consistent with the explicit non-locality of the entangled photons, on which this particular experiment is based on
    Therefore, if the Observer is outside space-time, there is no contradiction in a signal photon SP1 registering appropriately at $T_0$, based on the Observer's 'which path' information about its entangled idler photon IP1, available at $T_1$.
    Or, since there is no time dimension in non-locality, there is no contradiction
    The only contradiction regarding an *apparent retrocausality* is in the mind of the human observer looking at the recordings in space-time.

12. We define this non-local Observer as NCQO – (Nonlocal Conscious Quantum Observer).
13. That ties in directly to non-locality as a quantum phenomenon, especially for entangled photons.
14. So NCQO has to exist in the quantum dimensions outside space-time.
15. For the NCQO, $T_0$ and $T_1$ have no meaning
16. We will now redefine, for convenience and brevity: NCQO as **O**– the non-local conscious **Observer** outside of space-time, and
17. Lco as **o** - that is the local conscious **observer** that exists *in space-time*, and is an integral part of **O**.

18. So **O** and **o** are related, and together, they are the bridge between space-time *manifestation* and outside space-time *potential*.
19. Let us now look at the specific representation of the wave function and its collapse:
20. The wave function is represented by $\psi$
    a. $\psi$ has an imaginary number component and so cannot be directly observed in space-time
    b. $\psi$ is a mathematical function that represents the complex-valued probability amplitude of the quantum state, and the probabilities for the possible results of measurements made on the system can be derived from it.
    c. This non space-time probability representation can only be observed by **O**, which is also outside space-time
    d. In the Kim experiment, interference patterns occur at joint detection points with $D_1$ and $D_2$
    e. When the which-path information is 'erased' at $T_1$ when recorded at $D_1$ and $D_2$, then the information is not available anymore, and **O** allows it to continue
    f. The local consciousness part of the observer **o** then records this as an interference pattern.
    g. Keynote: The interference pattern that is recorded is only a space-time representation of the complex-valued probability amplitude of the quantum state, without collapse to a single space-time coordinate.
21. So it is not really a conventional 'wave', but an interference pattern that emerges from the space-time representation of the complex-valued probability amplitude of the quantum state.
22. Separately, we can do the math to show that this is what the interference patterns represent.
23. Simple diffraction/particle at joint detection rates with $D_3$ and $D_4$
    a. When the which path information of the idler photon at $D_3$ and $D_4$ is observed at $T_1$ and then mapped to the corresponding signal photon at $D_0$, at time $T_0$, then **O** observes this as something that can be represented in space time as a collapsed wave function, i.e. a particle with a unique representation in space-time.
    b. **O**, outside space-time, then communicates that it has this which path information, with value $\psi$ (with its imaginary number component outside space-time), to **o**, which is in space-time
    c. **o** recognizes this information by matching it with its own implicit equivalent complex conjugate of $\psi$
    d. This mutual 'recognition' and fit'(perhaps in the same way as chemical and biological bonds between matching molecules 'fit into' and latch on to each other) causes $\psi$ to get magnified by its complex conjugate – this is the mathematical meaning of 'squaring'.
    e. $\psi$ squared by its complex conjugate results in $|\psi|^2$. This is a real number that can be represented in space-time as a 'particle' appearing in specific space-time regions.
    f. The local consciousness part **o** then records this as a non-interfering simple diffraction pattern.

### Retrocausality

There are two implications for the phenomena of retrocausality:

The first, assumes that the Cause is the act of observation by a local conscious observer and the Effect is the manifestation as particle, i.e. the collapse of the wave function. Therefore, when the observation is made after the effect, retrocausality is a possible explanation. However, we have shown that observation by **o** is not the cause. **o** uses information from **O** to present the recording of the information available - this is not the effect.

The second implication is that Causality is only meaningful in space-time. **O** is outside space-time. **o** sees and records, in space-time, the then-available information. **O** is not subject to retrocausality, since it is non-local and outside space-time, hence cause and effect separated by time has no meaning. However, the appearance of retrocausality is only for the local (human) observer **o**, who exists in space-time.

If a photon in flight is interpreted as being in a so-called "superposition of states," i.e. if it is interpreted as something that has the potentiality to manifest as a particle or wave, but during its time in flight is neither, then there is no time paradox. This is the standard view, and recent experiments have supported it [13][14]. However, this does not consider the role of the observer and the implications of the act of observation as we have shown above.

In Wheeler's approach on Retrocausality, Wheeler states that
> "The thing that causes people to argue about when and how the photon learns that the experimental apparatus is in a certain configuration and then changes from wave to particle to fit the demands of the experiment's configuration is the assumption that a photon had some physical form before the astronomers observed it. Either it was a wave or a particle; either it went both ways around the galaxy or only one way. Actually, quantum phenomena are neither waves nor particles but are intrinsically undefined *until the moment they are measured*" (see also [15]).

We claim that this commonly held view, highlighted in italics above, is partially true and in specific experiments where there is clear measurement by a local observer. This would be the case in orthodox quantum mechanics if one did not have the additional complication of eraser situations, as described above. However, in the more general case, the 'definition' of the quantum phenomenon is not based on the *moment* (time) of *measurement*, but on the *availability of information both within and outside space-time*.

## CONCLUSIONS

Non-locality is an incontrovertible and significant quantum phenomenon [16][17] that presents a totally different paradigm than local, classical reality. The issues of retrocausation and non-locality were preliminarily discussed in the previous Retrocausation conference [18]. H.P. Stapp has advanced our understanding of quantum phenomena and the role of the mind [19]. In the Orthodox interpretation of quantum mechanics [19], which extends the Copenhagen Interpretation as achieved by J. von Neumann [20], the role of observation played by an observer and the response by Nature are outlined. Although we agree with the basic approach of the role of the mind in the Copenhagen Interpretation and its Orthodox refinement by von Neumann, we have shown here that the issues are more complicated when quantum eraser non-locality enters the picture as collapse occurs without an act of observation by a local observer; rather through these *apparent* retrocausal, quantum eraser experiments, which imply the availability of information outside of space-time and which are consistent with theoretically and experimentally established quantum non-locality [21]. The approach we take here is agnostic as to the role of Nature and possible modifications to standard Orthodox quantum mechanics [22]. The roles of **O** and **o** presented here are novel interpretations, consistent with the advanced eraser experiments of Kim *et al*. [12]. We actually believe that von Neumann [20] himself implied the existence of such an Observer as he concluded that the so-called 'Heisenberg Cut' was probably nowhere to be found. In our view that would be consistent with the existence of **O**. In future publication we will examine the full implications and how these points outlined here may be in agreement with non-physical views of reality [23] and related mathematical models [24].